\input harvmac
\input epsf.tex
\def \OO {{\cal O}}
\def \J {{\cal J}}

\def \De {{\Delta}}

\def \ra {\rightarrow}
\def \Lra  {\Longrightarrow}

\def \N {{\cal N}}

\def \e {{\eta}}

\def \del {\partial}

\def \a {\alpha}
\def \b {\beta}
\def \r {\rho}
\def \s {\sigma}
\def \p {\phi}
\def \m {\mu}
\def \n {\nu}

\def \t {\theta}
\def \Ga {{\Gamma}}
\def \td {\tilde }
\def \d {\delta}

\def \P {\Phi}

\def \ov {\over }

\def \fourth{{{1\over 4}}}

\def \inf {{\infty}}

\def \t {\vartheta}

\def \ss {\xi}

\def \KK {{\cal K}} 

\def \vex {{\vec{x}}}

\def \ha {{{1 \over 2}}}

\def \Box {{\del^2}}

\def \S {{\Sigma}}

\def \l {\lambda}

\def \lr { \lref}
\def\np {{  Nucl. Phys. }}
\def \pl {{  Phys. Lett. }}

\def \pr  {{ Phys. Rev. }}

\def \cmp {{ Commun. Math. Phys. }}

\newcount\figno
\figno=0
\def\fig#1#2#3{
\par\begingroup\parindent=0pt\leftskip=1cm\rightskip=1cm\parindent=0pt
\baselineskip=11pt
\global\advance\figno by 1
\midinsert
\epsfxsize=#3
\centerline{\epsfbox{#2}}
\vskip 0.2cm
{\bf Figure \the\figno:} #1\par
\endinsert\endgroup\par
}
\def\figlabel#1{\xdef#1{\the\figno}}
\def\encadremath#1{\vbox{\hrule\hbox{\vrule\kern8pt\vbox{\kern8pt
\hbox{$\displaystyle #1$}\kern8pt}
\kern8pt\vrule}\hrule}}

\baselineskip 15pt
\Title{\vbox
{\baselineskip 6pt
{\hbox {Imperial/TP/97-98/060}}
{\hbox{hep-th/9807097}} 
{\hbox{   }}
}}
{\vbox{
\centerline {On Four-point Functions}
\smallskip
\centerline { in the CFT/AdS Correspondence }
\vskip4pt }}

\centerline  { Hong Liu\footnote {$^*$} 
{e-mail address: hong.liu@ic.ac.uk} 
and 
A.A. Tseytlin\footnote{$^{\star}$}
{\baselineskip8pt e-mail address:
tseytlin@ic.ac.uk}\footnote{$^{\dagger}$}{\baselineskip8pt
Also at  Lebedev  Physics
Institute, Moscow.} 
}



\medskip

\medskip
\smallskip\smallskip
\centerline {\it  Theoretical Physics Group, Blackett Laboratory,}
\smallskip
\centerline {\it  Imperial College,  London SW7 2BZ, U.K.}
\bigskip\bigskip
\centerline {\bf Abstract}
\medskip
\baselineskip14pt
\noindent
\medskip
We discuss the properties of four-point functions 
in the context of the correspondence between a classical
supergravity 
theory in the bulk of the Anti de Sitter  (AdS) space 
and quantum conformal field theory (CFT) at the boundary.
 The contribution to a four-point function 
from the exchange of a scalar field of arbitrary mass in AdS space
is explicitly identified with that of the corresponding operator 
in the conformal partial  wave expansion of a four-point function on the CFT
side. Integral  representations are found for the  massless vector 
and graviton exchanges. We also discuss some aspects of 
the four-point functions of $tr F^2$ and $tr F F^*$ 
(`dilaton' and `axion') 
operators
 in the $\N =4$ supersymmetric  $SU(N)$ Yang-Mills  
 theory as  predicted by type IIB supergravity
in the $AdS_5$ background.

\Date {July 1998}

\noblackbox \baselineskip 18pt  


\lr \thooft{G. 't Hooft, ``A Planar diagram theory for strong interactions'',
Nucl. Phys. B72 (1974) 461.}

\lr \pol{ A.M. Polyakov, ``String theory and
quark confinement", hep-th/9711002; 
 ``A few projects in
string theory", Les Houches Summer  School,  1992:783, 
hep-th/9304146. }

\lr \mal{ J. Maldacena, ``The large $N$ limit of
superconformal
field theories and supergravity", {
hep-th/9711200}.}

\lr \gkp {S.S. Gubser, I.R.  Klebanov and A.M.  Polyakov,
``Gauge theory correlators from non-critical string theory",
hep-th/9802109.}

\lr \witt {E.  Witten, ``Anti de Sitter space and holography", 
hep-th/9802150.}

\lr \ferrz{ S. Ferrara,   C. Fronsdal and A. Zaffaroni, 
``On N=8 Supergravity on $AdS_5$ and $N=4$
Superconformal Yang-Mills theory", 
hep-th/9802203.}

\lr \ferza{ S. Ferrara  and A. Zaffaroni, 
``Bulk gauge fields in AdS supergravity and supersingletons",   
hep-th/9807090.}

\lr \ooo{ G. Horowitz and H. Ooguri,
``Spectrum of Large N Gauge Theory from Supergravity",
 Phys. Rev. Lett. 80 (1998) 4116, hep-th/9802116.
 }

\lr \kl { I.R. Klebanov, ``World-volume approach
to absorption by non-dilatonic branes",
\np B496 (1997) 231, hep-th/9702076;
S.S. Gubser, I.R.  Klebanov and A.A.
  Tseytlin, ``String theory and classical
  absorption by three-branes", 
  \np B499 (1997) 41, hep-th/9703040;
S.S. Gubser and I.R. Klebanov,
 ``Absorption by branes and Schwinger
terms in the world-volume
theory", Phys. Lett. B413 (1997) 41, 
hep-th/9708005.}

\lr\hashi{S.S. Gubser, A. Hashimoto, I.R. Klebanov  and  M.
       Krasnitz,
       ``Scalar absorption and the breaking of the world volume conformal
       invariance",
       hep-th/9803023;
 S. S. Gubser and A. Hashimoto, ``Exact absorption probabilities for the D3-brane'',
hep-th/9805140.}

\lr \volo{I.Ya. Aref'eva and  I.V. Volovich,
   ``On large $N$ conformal theories, field theories in Anti de Sitter 
space  and singletons", hep-th/9803028;
``On the Breaking of Conformal Symmetry in the AdS/CFT Correspondence", hep-th/9804182.}

\lr \free{D.Z. Freedman, S.D. Mathur, A. Matusis and L. Rastelli, 
``Correlation functions in the CFT$_d$/AdS$_{d+1}$
correspondence", hep-th/9804058.}

\lr\freed{D.Z. Freedman, S.D. Mathur, A. Matusis and L. Rastelli,
to appear,  talk by D.Z. Freedman
at Strings 98, Santa Barbara. }
 
\lr \liu{H. Liu and A. A. Tseytlin, ``D=4 Super-Yang-Mills, D=5 Gauged
Supergravity, and D=4 Conformal Supergravity,'' hep-th/9804083.}

\lr\muck{
W.  M\" uck and K.S. Viswanathan, 
``Conformal field theory correlators from classical scalar field theory on AdS$_{d+1}$", 
hep-th/9804035.}

\lr \muckv{W.  M\"uck and  K. S. Viswanathan, ``Conformal Field
Theory Correlators from Classical Field Theory on Anti-de Sitter Space II.
Vector and Spinor Fields,'' hep-th/9805145.}

\lr\nensfet{
M. Henningson and  K. Sfetsos,
``Spinors and the AdS/CFT
correspondence",  hep-th/9803251. }

\lr \rchalmers{G. Chalmers, H. Nastase, K. Schalm and R. Siebelink, 
 ``$R$-Current Correlators in $\N=4$ SYM from $AdS$,''
hep-th/9805015.}

\lr \ramir{A. M. Ghezelbash, K. Kaviani, S. Parvizi and  A. H. Fatollahi,
``Interacting Spinors-Scalars and the AdS/CFT Correspondence,''
hep-th/9805162.}

\lr \sei{S. Lee, S. Minwalla, M. Rangamani and N. Seiberg, 
``Three-Point Functions of Chiral Operators in D=4, $\N=4$ SYM at Large N'',
hep-th/9806074.}

\lr \fro{G.E. Arutyunov and S. Frolov, ``On the origin of supergravity boundary
terms in the AdS/CFT correspondence'', hep-th/9806216.} 

\lr \howe{P.S. Howe and P.C.  West, ``Operator product expansions in 
four-dimensional superconformal field theories'', 
Phys. Lett. B389 (1996) 273, hep-th/9607060.}

\lr \ansel{D. Anselmi, M. Grisaru and A. Johansen, ``A Critical Behaviour 
of Anomalous Currents, Electric-Magnetic Universality and $CFT_4$'', 
hep-th/9601023,
Nucl.Phys. B491 (1997) 221.}

\lr \allen{
B. Allen, ``Graviton propagator in de Sitter space'', \pr D 34 (1986) 3670;
B. Allen and M. Turyn, ``An evaluation of the graviton propagator
in de Sitter space'' \np B292 (1987) 813.}

\lr \fpf{E.S.  Fradkin and M.Ya. Palchik,
   ``Conformal Quantum Field Theory in D
   dimensions"  
   (Kluwer, Dordrecht, 1996).}

\lr \fps{E.S.  Fradkin and M.Ya. Palchik,
   Phys. Rept. 44C (1978) 249.}

\lr \tmp{I.T.  Todorov, M.C. Mintchev and V.B. Petkova
   ``Conformal Invariance in Quantum Field Theory 
   (Scuola Normale Superiore, Pisa, 1978).}

\lr \cpw{S. Ferrara, R. Gatto, A.F. Grillo and G. Parisi,
Nuovo Cim. Lett. 4 (1972) 115; 
Nuovo Cim. Lett. 5 (1972) 147; A. M. Polyakov, JETP 39 (1974) 10;
M. Ya. Palchik, \pl 66B (1977) 259; G. Mack \cmp 53 (1977) 155.}

\lr \ferr{S. Ferrara, R. Gatto, A.F. Grillo and G. Parisi,
\np B49 (1972) 77; 
Nuovo Cimento 26 (1975) 226.}

\lr \banks{T. Banks and M.B. Green, 
``Non-perturbative Effects in $AdS_5 \times S_5$ String Theory and d=4 SUSY 
Yang-Mills'', hep-th/9806004; M. Bianchi, M.B. Green, S. Kovacs and 
G. Rossi, ``Instantons in supersymmetric Yang-Mills and D-instantons 
in IIB superstring theory'', hep-th/9807033.}

\lr \jacob{B. Allen and T. Jacobson, ``Vector two-point functions
in maximally symmetric spaces'', Commun. Math. Phys. 103 (1986) 669.}

\lr\freedm{D.Z. Freedman, S.D. Mathur, A. Matusis and L. Rastelli, 
``Comments on 4-point functions in the CFT/AdS correspondence",
 hep-th/9808006.}

\newsec{Introduction}

There is a recent revival  of interest  in  the connections between 
large $N$ Yang-Mills theory \refs{\thooft} and  string theory \refs{\pol}  
following, in particular,  
the conjecture \refs{\mal} that there is an exact correspondence between 
string/M theory on  the Anti de Sitter space 
$AdS_{d+1}$  and certain superconformal field theory
(CFT$_{d}$) defined at  the boundary of the $AdS_{d+1}$
(see also  \kl).
 According to the conjecture, the quantum 
$\N=4$ supersymmetric Yang-Mills theory
 with gauge group $SU(N)$ in the large $N$
and 
large 't Hooft coupling limit 
can be described by the classical type IIB supergravity
 on  $AdS_5 \times S^5$ space.

The formulation  of the conjecture was  made more explicit 
 in \refs{\gkp,\witt},
where it was proposed that the partition function of supergravity/string theory 
with fixed boundary values of the fields 
 is to  be  identified with the generating functional 
of the  composite operators in the 
CFT. There is a one-to-one correspondence between 
certain   local operators $\P_i$  of the  boundary CFT and the bulk 
fields $\p_i$ in AdS \refs{\gkp,\ooo,\witt,\ferrz}. The boundary CFT operator $\P_i$ and  the associated 
bulk field $\p_i$
 carry the same unitary, 
irreducible and the highest weight representation of the 
conformal group $SO(d,2)$, where the scale dimension $\l_i$ of $\P_i$
is identified with the lowest energy 
value of $\p_i$  and can  be further 
related to the mass  of $\p_i$. 
The 
correlation functions of the CFT operators are identified with 
the classical ``S-matrix elements"
 of the  bulk fields with their 
boundary values fixed. Two-point and three-point functions 
follow simply from the quadratic terms and cubic vertices 
of the bulk theory, while  the four-point functions,    
in general,   contain the contact contributions as well as 
the exchanges of virtual particles.

Using this proposal, some   `model' and `realistic'  
two-point and three-point functions have been computed  in 
\refs{\volo, \muck, \nensfet,  \free, \liu, \rchalmers, \muckv, \ramir}. 
In  particular,  a family of three-point functions of chiral primary
operators in $\N=4$ SYM have been evaluated and shown to be equal to
their free-field values, suggesting a non-renormalization 
theorem in large $N$ limit \refs{\sei}.

In this paper we investigate the properties of the 4-point functions 
in the context of the  
CFT/AdS correspondence. Four-point functions from contact 
interactions (quartic vertices) 
were considered  before in \refs{\muck}. The 
scalar exchange diagrams with some special values of mass  were
also discussed in \refs{\freed}.
The exchange diagrams in AdS
are,  in general,  very difficult to evaluate 
explicitly, as the propagators 
and the integrals are  quite complicated. Here we  follow 
 a different 
approach.

In CFT$_d$, the states generated by acting  by a  product of 
the conformal operators
on the vacuum can be decomposed into a direct sum of irreducible representations 
of the conformal group
\eqn\wave{
\P_{1}(x_1) \P_{2}(x_2)|0> = 
\sum_k \int d^d x \, \, Q_{k}(x|x_1,x_2)
|k, x>\ ,
}
where $k$ sums over all the irreducible representations in the Hilbert space 
and states $|k, x> = \P_k (x) |0>$ span the space of irreducible   
representation $T_{k}$.
This  conformal partial wave expansion (CPWE) was 
obtained in early
seventies by several authors \refs{\ferr,\cpw} (see
\refs{ \fpf, \fps, \tmp}\ for 
 reviews). 
 Using  \wave, the  four-point functions can be written 
as
\eqn\cofour{
<0|\P_{1}(x_1) \P_{2}(x_2) \P_{3}(x_3) \P_{4}(x_4) |0> = 
\sum_k G_k\ , 
}
where $G_k$ is the contribution to  the four-point function from the intermediate
states generated by the  operator $\P_k$, 
\eqn\wcon{
G_k = \int d^d x d^d y \, Q^*_{k} (x_1, x_2|x) <k,x|k,y>
Q_{k} (y|x_3, x_4) \ . 
}
Since the bulk propagator of the field $\p_k$ in $AdS_{d+1}$ 
 associated to the  conformal operator $\P_k$ by the AdS/CFT correspondence can  be also  
written as a 
 sum over  normal modes in $AdS_{d+1}$ which  span the same   irreducible 
representation $T_k$ of the isometry/conformal  
group $SO(d,2)$, one is tempted 
to conjecture that  $G_k$  represents the contribution of the 
$\p_k$ exchange diagram
in $AdS_{d+1}$. 
Diagrammatically, this equivalence can be expressed as

\fig{
Equivalence between CPWE and scattering in AdS.
}{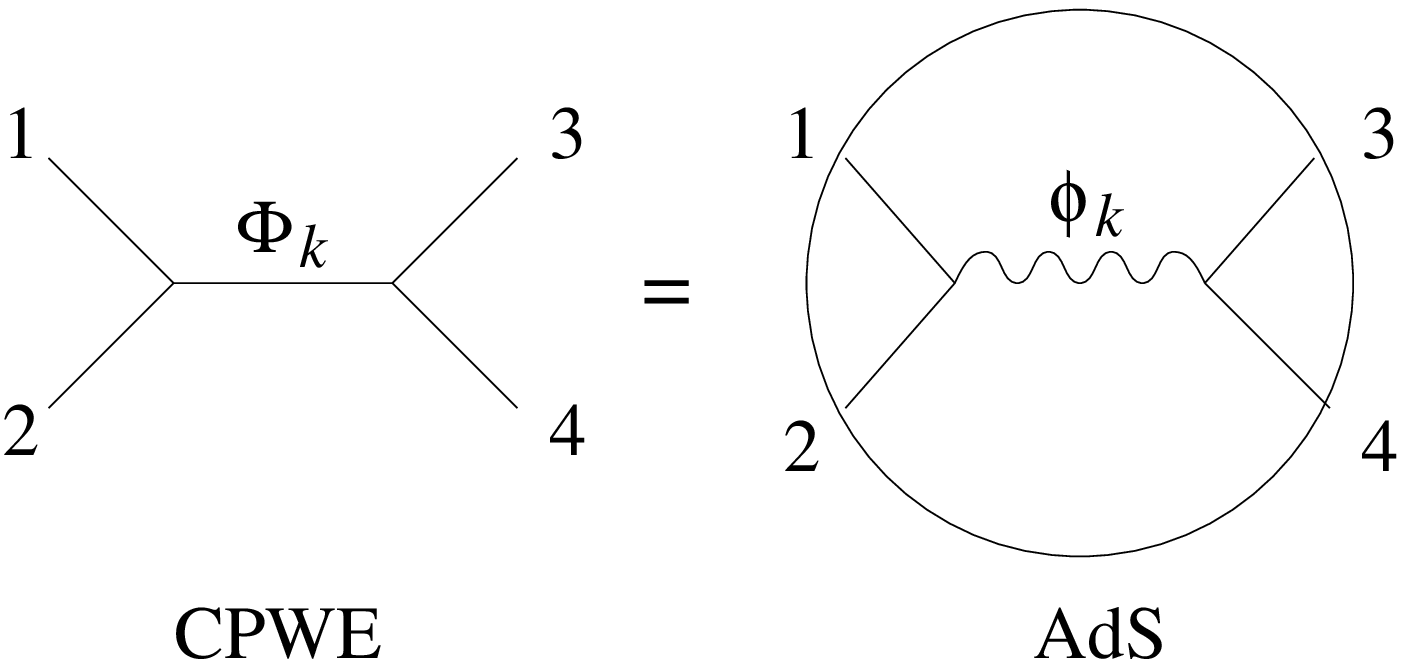}{6.8 truecm}
\figlabel\equi

One of the aims of the present  paper is 
 to prove    that this is indeed the case
 for the  intermediate states corresponding to  the 
scalar operators.
It should be  possible to generalise our  method 
 also  to  operators of higher spin. 
We shall   attempt to consider the 
 two cases which 
are of particular interest: massless vector and massless 
tensor (graviton)
exchanges.  They correspond 
to the conserved current vector and the energy momentum tensor 
operators in the CFT. 
The explicit demonstration of the equivalence 
between the   CPWE representation for the CFT correlator
 and the AdS amplitude here appears to be 
 more difficult and will not be given in the present paper.
 The  expressions for the 
propagators for the photon and graviton in AdS  are 
 quite 
involved (useful  expressions for them which are 
suitable for explicit calculations 
were not previously given  in the literature, cf. \refs{\jacob, \allen}). 
There are also complications 
related  to the presence 
of the gauge degrees of freedom and the fact 
  that the current and the energy momentum 
tensor carry indecomposable representations of the conformal 
group. In this paper we shall use  a non-covariant 
gauge fixing and  will be  able to 
write down the AdS amplitudes with the photon and graviton 
exchanges in the complete    integral form. 
The detailed analysis  of these amplitudes 
and establishing their relation to CPWE  will
not be attempted here.

Having identified the exchange diagrams with CPWE, a question that naturally arises 
is the interpretation of  contact interactions 
(quartic or higher vertices) on the CFT side,
as there is no obvious counterpart  for them in CPWE. 
One possibility is that since contact terms 
can  always  be formally written as  special  exchange 
diagrams,
e.g.,
 \eqn\voo{
\int d^{d+1} x \, \p_{1} \p_2 \p_3 \p_4 
= -\int d^{d+1} x\  d^{d+1} y \, \ \p_1 \p_2\  G(x,y)\   \del^2 ( \p_3   \p_4)
\ , }
where  $G$ is the massless field propagator, 
 $-\del^2 G = \d (x-y)$, 
they might be contained in  CPWE. 
Another possibility could be that the boundary CFT  is not closed.
Let us consider for example the $\N=4$ SYM theory 
in the  large $N$ and large `t Hooft coupling $\lambda=g^2 N$ 
limit. Suppose that a four-point function  at 
finite $\lambda$  is expanded in the form of \cofour\ and \wcon. 
As we increase  $\lambda$, certain 
correlators may  approach  zero as some inverse powers of $\lambda$ and thus 
may not  contribute 
 in the $\lambda \to \infty$ 
 limit. But if there is a very large number of such 
vertices, their  total  contribution to the sum \cofour\ may not vanish.
This  would   correspond to the presence of 
contact interactions
in 
supergravity.\foot{These contact supergravity vertices may be thought of as 
originating from string field theory  (with only cubic interactions 
between massless and massive modes) in the low-energy approximation 
in which all massive string  modes are integrated out.}
As there are many  contact terms in IIB supergravity on $AdS_5 \times
S^5$, 
such possibility  deserves a detailed  investigation.

The existence of the AdS/CFT correspondence puts   by itself 
 strong constraints on the theories on both sides. Since 
conformal field theories always  contain 
the energy-momentum tensor which generates the conformal algebra, 
 the theory in AdS must be a gravitational
theory. If the theory in AdS is a field theory (supergravity),  
then on CFT side the possible three-point functions and intermediate states contributing
to \cofour\ are highly constrained as only vertices which can be written
as local invariants in AdS are allowed.  For example, consider 
a  correlation function of four scalars.
In general,  symmetric tensor operators  of spin greater than 
two can contribute  to it as  intermediate states in \cofour.
However,  
 there is no local covariant  interaction vertex for two scalars
and a higher spin tensor in  supergravity theory (though it may be present in string theory),
so it should vanish also on the CFT side.
 Assuming  
the equivalence between CPWE in CFT and 
the scattering amplitude in AdS, 
we see also  that different channels for CPWE in CFT  
should correspond   
to  $s-t-u$ channels in the scattering amplitudes in
AdS, which seems to  imply  that the scattering 
amplitudes in AdS should have 
$s-t-u$ crossing  symmetry.\foot{One could think that this
 might  be an indication  
that the theory on the  AdS side should actually be 
a string-type theory. One does not expect to find 
the crossing symmetry  in the bulk 
supergravity amplitudes but  this is less clear when 
  the bulk-to-boundary propagators are attached. Duality    
 is, of course,   restored in the bulk amplitudes 
once one
replaces the supergravity  amplitudes by the full string amplitudes, i.e.
includes all $\alpha'$-corrections. At the same time, 
the  boundary theory at large $N$ and large $g^2N$ corresponding
just to supergravity with no $\a'$-corrections is also  a CFT 
which should have CPWE.}

In  case of the ``$\N=4$ SYM --  IIB supergravity on $AdS_5\times S^5$" 
correspondence
 the 
supergravity four-point functions in general are  quite
complicated.\foot{Higher-order $\a'$
and non-perturbative  string-theory
contributions to four-point functions in $\N=4$ large $N$ SYM 
were discussed  in \refs{\banks}.} 
In this paper we shall focus on
the dilaton-axion sector, where  
the corresponding four-point functions  are given  
by a  relatively small number of diagrams.
We will  show that 
  the  main  non-trivial contribution  in this sector
comes from a graviton exchange (the `mixed' scalar 
four-point function contains also a contact contribution).
 Using the graviton propagator 
found here  in the non-covariant 
$h_{0\m}=0$ gauge, we will be  able to  present  the complete
expressions for   the scalar 
four-point 
functions in terms of the formal integrals.  We shall
discuss briefly certain  
properties of the integrals, leaving their  
 evaluation and  
establishing the correspondence
with the  CPWE in CFT   for future.

The structure of the paper is as follows. In section 2 we shall review 
some  aspects of CFT in $d$ dimensions, in particular, 
 the conformal partial 
wave expansion for  the four-point functions. In section 3 
we shall 
 discuss the scattering diagrams involving scalar exchanges 
in AdS space. We will show that  
they can be identified with the contributions to CPWE coming from the corresponding 
operators  on the CFT side. In section 4  we
shall study the scattering amplitudes 
involving exchanges of massless vector and graviton. 
In section 5 we shall  consider the dilaton and axion 
  four-point
functions  in $D=5$ supergravity 
corresponding to the correlators of the 
 $tr F^2$  and $trF F^*$  operators 
in $\N=4$ SYM theory.
 Appendix A  contains notation and 
 some  technical details about the scalar 
propagator in $AdS_{d+1}$. 
In Appendix B  we recall the expression \ferr\ 
(see also \ferza)  for 
the scalar operator contribution to the scalar 
four-point function in CFT$_d$.

\newsec{Four-point functions and conformal partial wave expansion 
in CFT}

Let us first review certain aspects of the conformal field theory
in $d$ dimension \refs{\fpf,\fps,\tmp}. Denote the space of an irreducible 
representation $T_{\s}$ of the conformal group\foot{For the
 Minkowskian signature, the representations under consideration 
  are those 
of the infinite covering group of $SO(d,2)$ which are unitary and 
satisfy the spectrality condition: $p_0 >0,\  p^2 >0$.  
For the  Euclidean signature, 
they are the irreducible representations of $SO(d+1,1)$
which can be analytically  continued from the Minkowskian counterparts.}
as $M_{\s}$, \ $\s =(\l, \vec{s})$, \  where $\l$ is the conformal dimension
and $\vec{s}$ is a set of quantum numbers labelling the spin degrees
of freedom.  
We assume that the Hilbert space can be represented as a direct sum 
of spaces $M_{\s_i}$, i.e.
\eqn\hilbert{
{\cal H} = M_{\s_1} + M_{\s_2} + ... + M_{\s_i} + ... \, , \,\,\,\,\,\, \ \ \ \
\s_i = (\l_i, \vec{s}_i), \,\,\, \ i=1,2,...\ . 
}
Conformal fields $\P_{\s_i}(x)$ are defined as the operators which generate
spaces $M_{\s_i}$,
$$
M_{\s_i} =  \{ |\s_i, x> \,\, {\rm for \,\, all \,\,} x \}
= \{ \P_{\s_i}(x) |0> \,\,  {\rm for \,\, all \,\,} x  \}. 
$$
States of the type $\P_{\s_i} (x_1) \P_{\s_j} (x_2)|0>$ can 
be decomposed  in terms of the basis in \hilbert,
\eqn\cpwe{
\P_{\s_i}(x_1) \P_{\s_j}(x_2)|0> = 
\sum_k \int d^d x \, \, Q_{ijk}(x|x_1,x_2)
|\s_k, x> \ . 
}
This implies  the operator product 
expansion (OPE) 
\eqn\ope{
\P_{\s_i}(x_1) \P_{\s_j} (x_2) = \sum_k \int d^d x \, \, 
Q_{ijk}(x|x_1,x_2)
\Phi_{\s_k} (x)\ . 
}
The standard OPE in the nearby 
 points can be obtained from \ope\ by
expanding the integrand  in $y=x_1-x_2$,
\eqn\ttt{
\P_{\s_i}(y) \P_{\s_j}(0)|_{y \ra 0} \ \sim\  \sum_{k,m} A_{\s_k,m}(y) 
\Phi_{\s_k,m}(0)\ , }
where $A_{\s_k,m} \sim y^{-(\l_i + \l_j -\l_k -m)}$  and $\P_{\s_k,m}$ are $m$-th 
order derivatives of the  field $\P_{\s_k}$.

When $\P_{\s}$'s are orthogonal to each other, $Q$'s are just the 
amputated three-point functions,
\eqn\tyt{
Q_{ijk} (x|x_1,x_2) = \int \! d^d x'\   W^{-1}_{\s_{k}} (x-x')
<0|\P_{\s_k}(x') \P_{\s_i} (x_1) \P_{\s_j}(x_2)|0>
\ , }
with 
$
W_{\s}(x-x') = <0|\P_\s(x) \P_\s(x')|0> 
$
and its inverse $W^{-1}$ defined by,
\eqn\inve{
\int \! d^d x \, W_\s(x_1 -x) W^{-1}_\s (x-x_2) = I_+(x_1-x_2) \ , } 
\eqn\iii{ I_+(x) =
{1 \ov (2 \pi)^d} \int \! d^d p\  \t(p^0)\  \t (p^2)\  e^{ipx} \ .
} 
That $I_+$ (instead of the Dirac $\delta$-function)
 appears on the right side of 
\inve\ follows from the spectrality condition.

States involving higher-order products of $\P_{\s}$'s can be 
written in the basis \hilbert\ by repeatedly using \ope. 
The problem of solving the 
 theory thus becomes equivalent to finding  the spectrum 
and the couplings for the  infinite set of fields $\P_{\s_i}$.

Applying \ope\ to the  four-point functions we find 
$$
W_{ijkl}(x_1,x_2,x_3,x_4) = \ 
<0|\P_{\s_i}(x_1)\P_{\s_j}(x_2)\P_{\s_k}(x_3) \P_{\s_l}(x_4)|0>
$$
\eqn\parf{
= \sum_m \int \! d^d x d^d y \, \, Q_{ijm}(x_1,x_2|x)\ 
W_{\s_m}(x-y)\  Q_{mkl}(y|x_3,x_4)\ . 
}
We now switch to  the Euclidean signature. 
The conformal partial wave expansion 
of the four-point function in Euclidean region
takes the form\foot{We denote 
the Euclidean correlators by $G$ and omit some subscripts 
to simplify the notation.}\  ($\s = (\l, \vec{s})$) 
\eqn\euco{
G(x_1, x_2,x_3,x_4) = 
\sum_{\vec{s}} \int_C d \l \,\, \hat{G}_\s  (x_1,x_2,x_3,x_4)\ ,  
}
where the integral  is a contour
integral in the complex plane of the conformal dimension $\l$ 
(see below) and 
\eqn\hvfo{
\hat{G}_\s (x_1,x_2,x_3,x_4) = n(\s) 
\int  \! d^d x \  G_{ij\td{\s}}(x_1,x_2,x)
G_{kl \s}(x,x_3,x_4)\ .
}
Here
$$
G_{kl  \s} (x,x_3,x_4) = <\P_{\s}(x) \P_{k} (x_3) \P_{l}(x_4)>
\ , \ \ \ \ \   
G_{ij\td{\s}}(x_1,x_2,x) =  
<\P_{i} (x_1) \P_{j}(x_2) \P_{\td{\s}}(x) >
$$
are the Euclidean three-point functions with
$\td{\s} = (d -\l, \vec{s})$ and $n(\s)$ is some 
normalization constant.  The field 
$\P_{\td{\s}}$ is called  the conformal partner (or shadow operator)
of $\P_\s$, related 
to it by\foot{Note that there is no spectrality condition in 
Euclidean space  so that  
$\De_{\td{\s}}$ and  $\De_{\s}$  can be chosen to satisfy
$$
\int \! d^d x \, \De_\s(x_1 -x) \De^{-1}_{\td{\s}} (x-x_2) = 
\d (x_1-x_2)\ . 
$$
}
\eqn\sha{
\P_{\td{\s}} (x)= \int \! d^d y \, \De_{\td{\s}} (x-y)
\P_\s (y)
\ , \ \ \ \ \  
\De_{\td{\s}}(x-y) = <\P_{\td{\s}}(x) \P_{\td{\s}}(y)> 
\ . }
$\hat{G}_\s$ includes the  contributions from both
$\P_{\s}$ and its conformal partner $\P_{\td{\s}}$. The 
integration
contour
in \euco\ is chosen so that 
to select  only  the contribution from $\P_\s$.
One 
can decompose $\hat{G}_\s$ into the parts coming from  $\P_{\s}$ 
and $\P_{\td{\s}}$,
\eqn\decom{
\hat{G}_\s = G_\s + G_{\td{\s}}\ .
}
Since $G_\s$ and $ G_{\td{\s}}$ have different pole structure ($x_1-x_2 \ra 0$) 
$$
G_{ij\s} \sim {1 \ov |x_1-x_2|^{\l_i + \l_j -\l}} \ ,
\,\,\,\,\,\,\ \ \ \
G_{ij\td{\s}} 
\sim {1 \ov |x_1-x_2|^{\l_i + \l_j +\l -d} } \ , 
$$
the decomposition \decom\ is unique.
Using \decom\ instead of the  contour integral 
we can rewrite 
\euco\ as 
\eqn\caa{
G(x_1, x_2,x_3,x_4) = 
\sum_{\s_i} G_{\s_i}  (x_1,x_2,x_3,x_4)\ . 
}
Besides \parf\ and \euco, there are two other ways to write partial 
wave expansions:  in terms of ``$u$-channel" and ``$t$-channel". 
The equivalence of the  three channels is guaranteed
 by the associativity of the operator algebra \ope. 
This is usually  called the {\it crossing symmetry}
 of the four-point functions. 
As was already mentioned in the Introduction, the crossing symmetry of  the CFT four-point functions  should 
have  interesting implications for 
 the  structure of the corresponding 
scattering amplitudes in the  AdS space.

\newsec{Four-point functions in CFT/AdS correspondence: \ scalar  exchange}

In this section we consider  the contribution to a
four-point function of the  exchange of a scalar
field of an arbitrary mass. 
For definiteness, we shall study
the `model'  four-point functions of scalar fields 
and consider scattering in AdS resulting from 
 vertices 
of the type $\p \p_1 \p_2$ and $\p \p_3 \p_4$.
The scattering amplitude is given by (see Appendix A
for our notations)
\eqn\fourt{
S_\n (x_1,x_2,x_3,x_4)=\int \! {d u_0 d^d u \over u_0^{d+1}}
{d v_0 d^d v \over v_0^{d+1}} \ 
\KK_{\l_1} (u,x_1) \KK_{\l_2} (u,x_2)  G (u,v)
\KK_{\l_3} (v,x_3) \KK_{\l_4} (v,x_4)\ , 
}
where $ \KK_{\l} (u,x) $ is the boundary propagator \witt\ 
corresponding to a conformal field 
with dimension $\l$
\eqn\bpro{
\KK_{\l} (u,x)  = c_{\l} ({u_0 \over |u-x|^2})^{\l}\ , 
}
and 
$G(x,y)$ is the bulk propagator for a scalar field of mass  
$m$ (see Appendix A) \muck,
\eqn\spro{
G(x,y) =  (x_0 y_0)^{d/2} \int {d^d  {k} \over (2 \pi)^{d}} \, 
e^{i \vec{k} \cdot (\vec{x}-\vec{y})} I_\n (k x_0^<) K_\n (k x_0^>)\ .
}
Here $I$ and $K$ are the modified Bessel functions,
the parameter $\n$ is  related to the mass by
 $\n = \sqrt{m^2 + {1 \ov 4} d^2}$, and 
$x_0^<$($ x_0^>$) is the smaller (larger) number among
 $x_0$ and $y_0$. The amplitude 
\fourt\ with \spro\ inserted is not manifestly conformally invariant. 
Its conformal invariance can be seen by using an alternative representation
for  $G(x,y)$ \refs{\jacob}, 
\eqn\iui{
G(x,y) =r z^{- \l } F(\l,\n-\ha;2 \n +1, z^{-1})
\ , }
where $r$ is a normalization constant, $F$  is a hypergeometric function 
and $z={(x_0 + y_0)^2 + (\vex - \vec{y})^2 \ov 4 x_0 y_0}$. It is clear 
from \iui\ that $G(x,y)$ is invariant under transformation
$x \ra {x \ov |x|^2}$, $y \ra {y \ov |y|^2}$.

Let us first look at the ``pseudo-propagator'', given by
\eqn\pspro{
\hat{G}(x,y) = (x_0 y_0)^{d/2} \int {d^d  {k} \over (2 \pi)^{d}} 
 e^{i \vec{k} \cdot (\vec{x}-\vec{y})}
K_{\n} (k x_0) K_{\n} (k y_0)\ .
}
The value of \fourt\ with \pspro\ inserted instead of \spro\ will
be denoted $\hat{S}$. After the substitution of boundary propagators 
and \pspro\  it can be written as 
\eqn\afour{
\hat{S_\n} (x_1,x_2,x_3,x_4) = \int {d^d  {k} \over (2 \pi)^{d}} 
F^*_{12} (k;\vex_1, \vex_2) F_{34} (k;\vex_3,\vex_4)\ , 
}
where 
\eqn\defof{
F_{12} (k;\vex_1,\vex_2) = \int \! {d u_0 d^d u \over u_0^{d+1}} 
u_0^{{d \over 2}}  
e^{- i \vec{k} \cdot \vec{u}}\  K_{\n} (k u_0) \, \KK_{\l_1} (u,x_1) \KK_{\l_2} (u,x_2)  \ .
}
The Fourier transformation for the  boundary propagator with
$\l = \n + {d \ov 2}$ is 
\eqn\fouri{
\KK_{\l}(u,x) = c_\l({u_0 \over |u-x|^2})^{\l} = \int {d^d k \ov (2 \pi)^d } \ 
e^{ i \vec{k} \cdot (\vec{u}-\vex)} f(k, u_0)\ , 
}
with 
$$
f(k, u_0) = { 1 \ov b_{\l}} u_0^{{d \ov 2}} k^\n K_{\n} (k u_0) \ , 
\ \ \ \ \ \   b_{\l} ={ 2^{\n -1} \Ga(\n)}\ . 
$$
Thus
$$
 u_0^{{d \over 2}} K_{\n} (k u_0)=  b_\l k^{-\n} \int \! d^d x \,\,  
e^{- i \vec{k} \cdot (\vec{x} - \vec{u})} \KK_{\l}(u,x) \ . 
$$
Plugging this into \defof, we find
$$
F_{12} (k;\vex_1,\vex_2) = {b_{\l}} k^{-\n} 
\int \! {d u_0 d^d u \over u_0^{d+1}}
\int \! d^d x \,\, e^{- i \vec{k} \cdot \vex}
\KK_{\l}(u,x) \KK_{\l_1} (u,x_1) \KK_{\l_2} (u,x_2)  
$$
$$
=  {b_{\l}} k^{-\n} \int \! d^d x \,\, e^{- i \vec{k} \cdot \vex}
G_{\l \l_1 \l_2} (x, x_1,x_2)
=  {b_{\l}} k^{-\n} G_{\l \l_1 \l_2} (k, x_1,x_2)\ ,  
$$
where 
\eqn\threep{
G_{\l \l_1 \l_2 } (x, x_1,x_2) = 
\int \! {d u_0 d^d u \over u_0^{d+1}} 
\KK_{\l}(u,x) \KK_{\l_1} (u,x_1) \KK_{\l_2} (u,x_2)  
}
is the three-point function according to 
  the CFT/AdS correspondence.
Similarly, we find
$$
F_{34} (k;\vex_3,\vex_4) = {b_{\l}}  k^{-\n} G_{\l \l_3 \l_4} (k, x_3,x_4)\ . 
$$
Then it follows from \afour\ that
\eqn\foume{
\hat{S}_\n = b_{\l}^2   \int \! {d^d  {k} \over (2 \pi)^{d}} \, 
G^*_{\l \l_1 \l_2} (k, x_1,x_2)
k^{-2 \n}  G_{\l \l_3 \l_4} (k, x_3,x_4) \ . 
} From \fouri\ and \threep\ we can see that
\eqn\shad{
G_{\td{\l} \l_1 \l_2} (k, x_1,x_2) = {b_\l \ov b_{\td{\l}}} k^{-2 \n}
G_{\l \l_1 \l_2} (k, x_1,x_2)
}
Thus \foume\ can be written as
$$
\hat{S}_\n = b_{\l} b_{\td{\l}}   \int \! {d^d  {k} \over (2 \pi)^{d}} \, 
G^*_{\td{\l} \l_1 \l_2} (k, x_1,x_2)
G_{\l \l_3 \l_4} (k, x_3,x_4) \ , 
$$
and  in coordinate space it becomes,
\eqn\coor{
\hat{S}_\n =  b_{\l} b_{\td{\l}}  \int \! d^d x \, \ 
G_{\l_1 \l_2 \td{\l}} (x_1, x_2,x) 
G_{\l \l_3 \l_4} (x, x_3,x_4)
\ . }
We notice that the above expression \coor\ for $\hat{S}_\n$ is precisely 
the same as the CFT expression 
\hvfo\ with $\s=(\l,0)$. Thus we have identified  the 
amplitude  $\hat{S}_\n$ with the CFT correlator 
$\hat{G}_\s$ in \euco.

Let us now  look at the relation between $S_\n$ \fourt\ 
and $\hat{S}_\n$ \afour. 
Using that 
$$
K_{\n} = {\pi \ov 2} {1 \ov \sin \n \pi} (I_{-\n} - I_\n)\ , 
$$
and 
$$
\int_0^\infty \! d x_0 \int_0^\infty \! d y_0 
= \int_0^\infty \! d x_0 \int_0^{x_0} \! d y_0
+  \int_0^\infty \! d y_0 \int_0^{y_0} \! d x_0\ , 
$$
it is easy to see that 
\eqn\adecom{
\hat{S}_\n = S_{\n} + S_{-\n} \ ,  
}
which can be understood as the sum of the 
 contributions from the fields of dimensions 
$\l$ and $d-\l$ respectively. 
Comparing \adecom\ with \decom\ and assuming that ${S}_\n$
has analytic dependence on $\n$, we find that with 
$\hat{S}_\n$ identified with $\hat{G}_\s$
in \hvfo, $S_{\n}$ and $S_{-\n}$ are equal  
to  $G_{\s}$ and 
$G_{\td{\s}}$, demonstrating  the required
 relation between 
the AdS amplitude $S_\nu$ 
 and the   CFT correlator.

It is easy to see that the above procedure applies without change 
to other types of 3-point interaction 
 (e.g., $\p \del \p_1 \del \p_2$) and to 
scattering of higher spin fields involving { scalar}
 exchange.
Thus we have established the correspondence between the exchange diagrams  
in AdS and the conformal partial wave expansion in 
the  CFT for the case 
of the {\it scalar}  intermediate states.

We mention here that the contribution of a scalar operator 
to the CPWE (eqs \parf, \euco) of a four-point function 
has been evaluated long time  ago in \ferr.  
The expression can be written in a closed 
form in terms of double hypergeometric functions (see Appendix B). From 
the identification of \fourt\ and \parf, we see that 
it can also be interpreted as the scattering amplitude \fourt\ in AdS.

\newsec{Photon and graviton propagators in 
$AdS_{d+1}$}

In this section we shall consider  the scattering amplitudes in AdS
involving massless vector and graviton exchanges. The photon and graviton 
propagators in covariant gauges 
were discussed before in \refs{\allen}. The expressions found  
were   complicated and do not seem to be useful  in 
explicit calculations. 
Here we shall choose
 the non-covariant gauges, $A_0 =0$ for the 
 vector and $h_{0 \m}=0$ for the  graviton. 
It turns out that the resulting  AdS propagators  are 
quite simple  
and  have structure similar to that of their
 flat space counterparts.

\subsec{Massless vector }

Let us start with  the case of a  vector field 
in  AdS space described by the action
\eqn\veee{
I =  \int d \a \, (\fourth F_{\m \n} F^{\m \n}
+ A_\m \J^\m)
\ , \ \ \ \ \ \ \ \   
\int d \a = \int \! d^{d+1} x \sqrt{g_0}\ . }
We fix the  Coulomb gauge $A_0 =0$. In this gauge 
the equation for $A_i$ ($i=1,...,d$) becomes
\eqn\vecte{
(\del_0^2 + \del_j^2) A_i - \del_j \del_i A_j -
{d-3 \ov x_0} \del_0 A_i = -{1 \ov x_0^2} \J_i\ . 
}
The equation for $A_0$ gives the constraint 
\eqn\comaz{
\del_0 \del_i A_i = {1 \ov x_0^2} \J_0\ . 
}
We decompose $A_i$ as
$$
A_i = A^\perp_i + \del_i \xi \ , 
$$
where $A^\perp_i$ is the  transverse part  and 
$\ss = {1 \ov \Box} \del_{i} A_i $ with $\Box = \del_i\del_i$.
Then  \vecte\ and \comaz\ reduce to
$$
\del_0 \ss = {1 \ov x_0^2}{1 \ov \Box} \J_0, \,\,\,\,\,\, \ \ \ \ \ \
(\del_0^2 + \Box) A^\perp_i - 
{d-3 \ov x_0} \del_0 A^\perp_i = -{1 \ov x_0^2} \J^\perp_i\ , 
$$
where $\J^\perp_i$ is the  transverse 
part of $\J_i$, i.e. $\del_i \J^\perp_i=0$.
Thus  the equation for  $x_0 A^\perp_i$ reduces to that for
a free scalar of $m^2=-(d-1)$ in $AdS_{d+1}$ 
space with a source $x_0 \J^\perp_i$.
 It can be easily 
solved to obtain
\eqn\sola{
x_0 A^\perp_i (x) =  \int \! d \a \, \ G(x,y)\  y_0 \J^\perp_i(y)\ , 
}
where $G(x,y)$ the propagator for a free scalar  with 
 $m^2=-(d-1)$.

Then 
\eqn\exfi{
I = \ha \int \!  d \a \, A_i \J^i = \ha \int \! d \a \,
(A^\perp_i  \J^i + \del_i \ss \J^i ) \ . 
}
Note that the conservation of the current gives
$$
D_\m \J^\m = 0 \,\, \Lra \,\, \del_i \J^i = - {1 \ov \sqrt{g_0}}
\del_0 (\sqrt{g_0} \J^0) \ . 
$$
After a partial integration, 
 use of the  current conservation and the 
substitution of \sola\ into \exfi\  we get, 
\eqn\xxx{
I = \ha \int \! {d x_0 d \vex \ov x_0^{d+1}}\,{d y_0 d \vec{y} \ov y_0^{d+1}}
\, \ x _0 y_0 \ \J^\perp_i (x) \ G(x,y)\  \J^\perp_i(y)
-  \ha \int \! {d x_0 d \vex \ov x_0^{d+1}}  \ \J_0 {1 \ov \Box} \J_0 \ . }
This expression determines the photon propagator.\foot{In \xxx\
there is also a boundary term 
$- \ha \int_{\del M} \! d^d x \; x_0^{1-d} \; \ss \J_0$. 
In the non-covariant gauge we are choosing, this term  is responsible 
for the Ward identity of three-point functions, but we do not expect it 
to contribute to higher-point functions.}

\subsec{Graviton }
We consider the metric as a sum of the AdS metric and a graviton perturbation
$g_{\m \n} =  {1 \ov x_0^2} (\d_{\m \n} + h_{\m \n})$
and consider the action (see also \liu)
$$
I = \int \! d \a \, (R - 2 \Lambda + \ha h_{\m \n} T^\m_\n )\ .
$$
In the following  we shall assume that 
there is no index raising  
for $h$, but for the 
energy-momentum tensor $T_{\m \n}$ the indices
will be  raised  by $g^{ \m \n}_0 = { x_0^2} \d^{\m \n}$.
We shall consider the gauge $h_{0 \m} =0$.
In this gauge, the linearised Einstein equations for $h_{ij}$ become (see also \fro)
$$
\del_0^2 (h_{ij} -\d_{ij} h) 
- {d-1 \over x_0} \del_0 (h_{ij} -\d_{ij} h) 
$$
\eqn\einij{
+ \ \del_{k}^2 \tilde{h}_{ij} -\del_l \del_i \tilde{h}_{jl}
 -\del_l \del_j \tilde{h}_{il} + \d_{ij} \del_l \del_m \tilde{h}_{lm}
= -2 T_{ij} \ , 
}
where  $h= \d_{ij} h_{ij} $ and $\tilde{h}_{ij} = h_{ij} - \ha
\d_{ij} h$. 
There are also  two  constraints  following from the 
$00$ and $0i$ components of the Einstein equations,
\eqn\einoi{
 \del_0 (\del_j h_{ij} - \del_i h) = 2T_{0i}\ , 
}
\eqn\einoo{
-  \del_i \del_j h_{ij} + \del_i^2 h - {d-1 \ov x_0} \del_0 h 
= 2T_{00}\ . 
}
We decompose $h_{ij}$ as
$$
h_{ij} = \bar h^\perp_{ij} + \del_i B^\perp_j + \del_j B^\perp_i 
+ \del_i \del_j \eta +
{1 \ov d-1} (\d_{ij} - {\del_i \del_j \ov \del^2}) h' \ , 
$$
with $\del^2 = \del_i \del_i$ and
$$
B^\perp_i = {1 \ov \Box} \del_k h_{ik} 
- {\del_i \ov (\Box)^2} \del_k \del_l h_{kl}\ ,
\,\,\,\,\,\,\,\   \eta = {1 \ov (\Box)^2} \del_k \del_l h_{kl}\ ,\ 
\,\,\,\,\,\,\, h' = h - {1 \ov \Box} \del_k \del_l h_{kl}\ . 
$$
It is easy to check that
$$
\d_{ij} \bar  h^\perp_{ij} =0\ , \ \ \ \,\,\,\,\,\,\, 
\del_i \bar h^\perp_{ij} =0\ ,  \,\,\,\,\,\,\, \ \ \ \
\del_i B^\perp_i =0 \ . 
$$
Then the  equations \einij--\einoi\  become 
\eqn\gratt{
(\del_0^2 + \del_k^2)\bar  h^\perp_{ij} - {d-1 \ov x_0} \del_0 \bar h^\perp_{ij}
 = -2 t_{ij} \ ,  \ \ \ \  t_{ij} \equiv P_{ijkl}T_{kl} \ , 
}
\eqn\consi{
\del_0 B^\perp_i = {1 \ov \Box} (2 T_{0i} + \del_0 \del_i h')\ , \,
\ \ \ \ \ 
\del_0 h' = - {2 \ov \Box} \del_j  T_{0j}\ , 
}
\eqn\conso{
\del_0 h = -{2 x_0 \ov d-1} T_{00} + {x_0 \ov d-1} \Box h'\ , 
}
where   $P_{ijkl}$ is the transverse  traceless projector
in flat space.

Thus  the equations  for the transverse traceless 
part  $\bar h^\perp_{ij}= P_{ijkl} h_{kl}$ reduce to that of  
a free massless scalar,  so that 
 we get ($d \b = d^{d+1} y \sqrt{g_0(y)}\ $)
\eqn\yyt{
h^\perp_{ij} (x)  = 2 \int \! d \b  \, \ y_0^2\  G(x,y)\  t_{ij} (y)
\ . }
Using the conservation of the energy-momentum tensor 
and the equations \consi\ and \conso\ we find that
the quadratic (graviton propagator) part of the   Einstein 
action takes the form
$$
I = \ha \int \! d \a\  h_{ij} T^{i}_j = 
\int d \a d \b  \,\, (x_0 y_0)^2\  t_{ij} (x)\ G (x,y)\  t_{ij}(y) 
- 2 \int d \a \ x_0^2 \ T_{0i} {1 \ov \Box} T_{0i} $$ 
\eqn\scag{
-\ 
{ 2 \ov d-1} \int d \a \ x_0^2 \ \del_i T_{0i} {1 \ov \Box} T_{00}
- {d-2 \ov d-1} \int d \a \ x_0^2 \ \del_i T_{0i} ({1 \ov \Box})^2  
 \del_i T_{0i} \ , }
where we have ignored boundary terms resulting from partial integration.

\newsec{ Scalar four-point functions
in $AdS_5$ supergravity / $\N =4$ SYM theory }

In this section we shall consider  the four-point functions 
involving  the scalar 
operators $\OO = tr F^2$ and $\OO' = tr F F^*$ in $\N=4$ SYM theory 
in the large $N$ and large `t Hooft coupling limit
as  predicted by the type IIB supergravity in $AdS_5\times S^5$. 
$\OO$ and $\OO'$  
correspond to the massless fields $\p$ and $C$ in $AdS_5$ 
coming from the dilaton and axion of type  IIB supergravity on
$AdS_5 \times S^5$. 
The four-point functions of interest are
$<\OO(x_1) \OO(x_2)  \OO(x_3)   \OO(x_4)>$,\
 $<\OO'(x_1) \OO'(x_2)  
\OO(x_3)   \OO(x_4)>$ and  $<\OO'(x_1) \OO'(x_2)  
\OO'(x_3)   \OO'(x_4)>$. They correspond  
in $AdS_5$ to the  tree-level scattering amplitudes 
 of four dilatons,  two dilatons 
and two axions, and four axions respectively. 
Thus we need to look for vertices
in the Lagrangian of  IIB supergravity on $AdS_5 \times S^5$
involving two dilatons (axions) and one other field
or quartic vertices involving only dilatons and axions.
For this purpose it is sufficient to concentrate on  the 
graviton-dilaton-axion 
sector of the  supergravity\foot{ Our notation will be as follows.
All fields in $D=10$ will be  written with hats on them  and their  indices. 
$\m, \n, ...$  will refer to the indices  of  $AdS_5$, while 
$\a, \b, ...$  will refer to those  of  $S^5$.} 
\eqn\twob{
\hat{I} = {1 \ov 2 \kappa^2_{10}} \int_{AdS_5} \! d^{5} x \int_{S^5} \! 
d^{5} y \, \sqrt{- \hat{g}}\  
\bigg[\hat{R} - \ha (\del \hat{\p})^2 
- \ha e^{2 \hat{\p}} (\del \hat{C})^2 \bigg]\ . 
}
Since the fields $\p$ and $C$ are associated with the 
zeroth spherical harmonics of $\hat{\p}$ and $\hat{C}$ on $S^5$, it is
clear that the desired  vertices cannot involve fields from 
higher spherical harmonics. Thus the  fields in \twob\  can be 
assumed to have 
only  $x$ dependence, i.e.
$$
\hat{\p}(x,y) = \p(x), \,\,\,\,\,\,\ \ \  
\hat{C}(x,y) = C(x), \,\,\,\,\,\, \ \ \
\hat{h}_{\hat{\m} \hat{\n}} (x,y) = \{h_{\m \n}(x), h_{\m \a}(x),
h_{\a \b} (x)\} \ . 
$$
After dimensional reduction to $AdS_5$ and 
a Weyl scaling\foot{The  resulting graviton $h'_{\m \n}$  differs
from $h_{\m \n}$ by a Weyl scaling, i.e. 
$h'_{\m \n} = h_{\m \n} + {1 \ov 3} 
g_{0 \m \n} h^{\a}_\a$, where $g_{0 \m \n}$ is the background metric in 
$AdS_5$.}
 the relevant part of \twob\ becomes
\eqn\diax{
I = {1 \ov 2 \kappa^2} \int_{AdS_5} \! d^{5} x \sqrt{- g'}\ 
\bigg[R - \ha (\del \p)^2 - \ha e^{2 \p} (\del C)^2 \bigg]\ , 
}
where $R =R (h')$ is the 5d Ricci scalar.
The   $\p-C$ sector of the IIB supergravity on $AdS_5 \times S^5$
is  thus very  simple. The only cubic vertex involving 
two $\p$ is $\del \p\del  \p\ h'$.
In particular, there is no 
$h^\a_\a  \p \p $ vertex, where $h^\a_\a$ is the trace of the internal 
 ($S^5$)  part of the metric (massive fixed scalar)
 which in the $\N=4$ SYM  theory corresponds to the
operator of the structure 
$\OO_8
= tr F^4 - {1 \ov 4} tr (F^2)^2 $.
Since  in  the  free  Maxwell theory there is a non-vanishing
three-point function  $<\OO \OO \OO_8>$ \refs{\hashi}, 
 this 
three-point function must have non-trivial dependence on the 
 `t Hooft coupling (cf. \sei).

Following the same reasoning, we can see that 
the  diagrams  contributing to the four-graviton scattering 
may come only  from  the 
$R$  term in $\diax$. In particular, 
there is no cubic vertex involving two gravitons and  
other
fields. This is consistent with the expectation \refs{\howe} that the 
OPE of the energy momentum tensor
 in $\N=4$ SYM theory closes on itself.\foot{This is 
not so in  most of four-dimensional theories. For a discussion 
of the case of the 
$\N=1$ supersymmetric theories see \refs{\ansel}.}

\fig{Scattering diagrams for $\p \p \p \p$ and $CCCC$.
Only $s$-channel diagrams are displayed here.
}{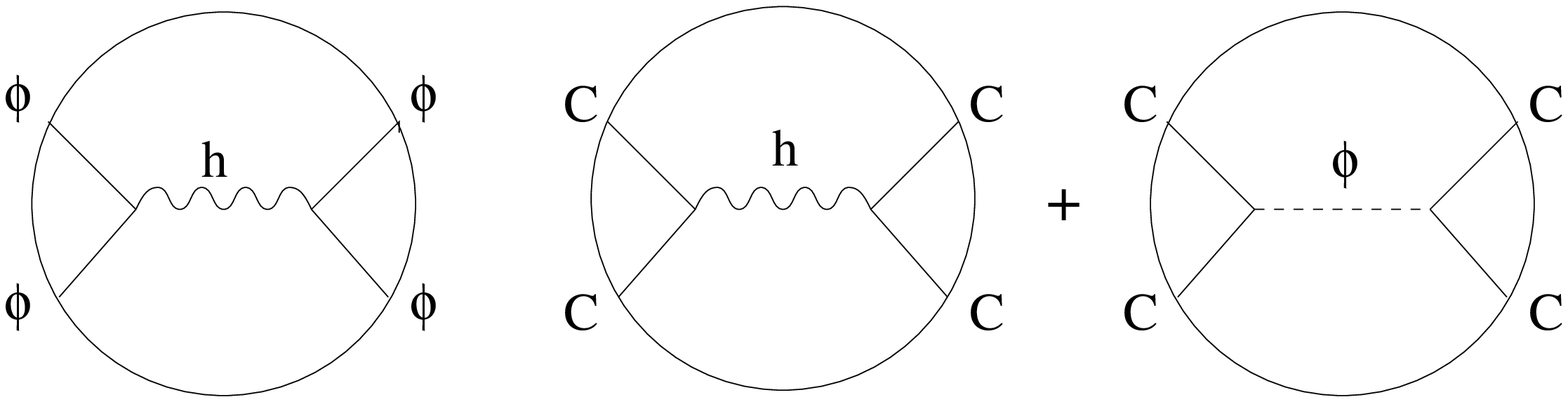}{10 truecm}
\figlabel\scatd

\fig{Scattering diagrams for $\p \p C C$.
}{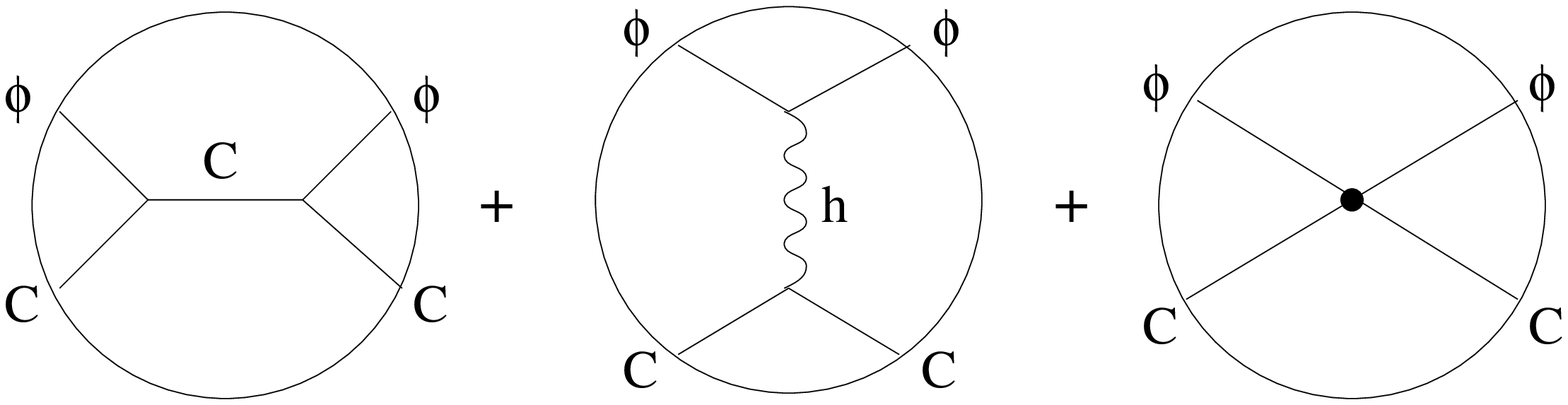}{10 truecm}
\figlabel\scatc

Starting with  \diax, we can write down the 
scattering diagrams in $AdS_5$ contributing to the  
four-point functions under consideration (see Fig.\scatd\
and Fig.\scatc). Let us first look
at the diagrams which do not involve graviton exchange. 
The scattering amplitude  with the   dilaton
or axion exchange, e.g., $\phi(1)C(2)\phi(3)C(4)$ in Fig.\scatc,  
can be written as
\eqn\iiu{
A_1 = \int \! {d u_0 d^d u \over u_0^{d+1}}
{d v_0 d^d v \over v_0^{d+1}} (u_0 v_0)^2
\del_{\m} \KK_{d} (u,x_1) \del_\m \KK_{d} (u,x_2)  G (u,v)
\del_{\n} \KK_{d} (v,x_3) \del_{\n} \KK_{d} (v,x_4)
\ . }
The contact $\p C \p C $ interaction   gives
\eqn\iuo{
A_2 = \int \! {d u_0 d^d u \over u_0^{d+1}}
\, u_0^2 \,
\KK_{d} (u,x_1)  \del_{\n} \KK_{d} (u,x_2) 
 \KK_{d} (u,x_3) \del_{\n} \KK_{d} (u,x_4)
\ .  }
Here  $\KK_{d}$ ($d=4$) and  $G(u,v)$ (see \bpro\ and \spro) 
are the boundary 
and bulk propagators for the massless scalars.
Since these  propagators satisfy 
$$
{1 \ov \sqrt{g_0}} \del_\m (\sqrt{g_0}g^{\m\n}_0 \del_\n) \KK_d = 0\ , \,\,\,\,\,\,\ \ \ \ \ 
{1 \ov \sqrt{g_0}} \del_\m (\sqrt{g_0}g^{\m\n}_0 \del_\n) 
 G(u,v) = - {1 \ov \sqrt{g_0}} \d (u-v)\  ,
$$ 
we find by a partial integration\foot{Boundary terms resulting from 
partial integration give  only unimportant contact terms (terms 
containing Dirac delta-function and its derivatives).}   
 that $A_1 (1,2,3,4)$ reduces to $-\ha A_2(1,3,2,4)$.

Direct evaluation of $A_2$ is quite tedious.
It is easy to show, however, that the symmetric
in $x_1,...,x_4$
part of \iuo\  $(A_2)_{sym}$ 
 (and thus  $(A_1)_{sym}$)  vanishes. 
 One way to see that is to   make a field redefinition,
 $\p = - \log (1 + \psi)$. 
Then the free action for $\p$ 
becomes
\eqn\fiedr{
I =   \int d^{d+1} x  \sqrt {g_0} \ (1 + \psi)^{-2} (\del \psi)^2
\ . }
The new  action  generates cubic and quartic dilaton vertices
and it can be seen that the resulting four-dilaton scattering amplitude  
is  proportional to  $(A_2)_{sym}$. Since  the 
field redefinition 
 changes the correlators only by unimportant  contact terms, 
we conclude 
that  $(A_2)_{sym}$  should contain
 only   contact terms, i.e. should  vanish for separated points.\foot{In the original version  of this paper we  were assuming    that this argument 
demonstrates the triviality of $A_2$, but it actually applies only to 
$(A_2)_{sym}$  as the the four legs  here correspond to the {\it same} field
(while $\psi^2 (\del \psi)^2$ reduces  after integration by parts 
to   $- {1 \ov 3} \psi^3 \del^2 \psi $
that  vanishes on shell, 
 there is 
no similar rearrangement for $\phi^2 (\del C)^2$).
We  noticed this mistake after  reading  the recent 
paper \freedm\
with conclusions of which we now agree.}

The fact that $(A_2)_{sym} = -2 (A_1)_{sym}=0$
 implies, in particular,  that as  in the case of   the $\p\p\p\p$ amplitude, 
the  only non-trivial 
contribution  to the $CCCC$ amplitude 
comes from the sum of the $s,t,u$ graviton 
exchanges, so that the $\p\p\p\p$ and $CCCC$
 amplitudes are actually  equal.\foot{This may be viewed 
as a consequence of the $SL(2,R)$ symmetry
of the action \diax.} 
The structure of the $\phi C \phi C$ amplitude is more complicated
 as  in addition to the  graviton exchange it contains also
the  sum of the axion  exchange and the  contact 
contributions  given by a combination of  
$A_2$ amplitudes
with  different orders of end-points (see also \freedm).

The  formal  integral expression for the
graviton contribution to the  four-point functions can be written 
down by substituting into \scag\ the 
appropriate expression for the 
energy-momentum tensor  corresponding to  
the massless scalar fields,
$$ 
T_{\m \n} = \del_\m \KK_d (u,x_1) \, 
\del_\n \KK_d (u,x_2) - \ha \d_{\m \n} \, \del_\l \KK_d (u,x_1) \,
\del_\l \KK_d (u,x_2)\ .
$$ 
Here  $\KK_d$  is again the   boundary propagator (with $d=4$).
The  resulting  expression is quite long  and  will
not be explicitly presented here. We leave its detailed  analysis 
for future work.

To conclude, let us  make  some  speculative  
remarks on consequences
of  possible    crossing symmetry of  CFT 
four-point functions   for the 
structure of  scattering amplitudes in  AdS.
 From  correspondence 
between the scattering amplitudes 
in AdS and  correlators  in CFT given by CPWE's, we would expect that  the  amplitudes
in Fig.\scatd\ and Fig.\scatc\  should possess $s-t-u$ symmetry.
That would imply, for example,  that the $s,t$ and $u$ channel  graviton exchanges
 in the $\p\p\p\p$ amplitude 
 are  all equal. Then in the $CCCC$ amplitude
(which does not contain contact contribution)  
the $s,t,u$ dilaton exchages would also need to be equal. 
Since their sum $(A_1)_{sym}$ vanishes, 
each of them should also be vanishing,  
at least up to  some  singular 
 contributions (cf. \freedm) 
that  would need to  be
 subtracted  as part of  a 
definition of  the CFT/AdS correspondence.\foot{Note that in Euclidean space
the  massless scalar 
 exchange diagrams are generated by a contact-type 
($\delta$-function) vertex $<\OO(x) \OO'(y) \OO'(z)> \sim
[\d (x-y) + \d (x-z)] {1 \over |y-z|^8} - \d (y-z) {1 \over |x-z|^8}$
 represented  in the bulk  by the  $\p \del C \del C$ interaction
(its non-contact part vanishes since  upon integration by parts it 
  reduces to $  \ha  C^2 \del^2 \p  - \p C \del^2 C$). 
However,  the corresponding Minkowskian (Wightman)  three-point 
function is zero and thus 
 should  not contribute to the  CPWE  
of the  four-point Wightman function in CFT.} 
To be able to draw any conclusions from the assumption 
of crossing symmetry 
in the case of the 
 $\p C \p C$ amplitude (Fig.\scatc) 
one  needs first  to 
interpret  the contact bulk   diagram contributing to it.
For example,  we may replace it by  a
 massless scalar exchange   in the  $t$-channel  using that, 
 as shown above, 
$A_2 (1,2,3,4)=- 2 A_2(1,3,2,4)$.\foot{The massless scalar
here  may be identified with  dilaton, as there 
is a non-vanishing contact-type
Euclidean three-point function in SYM theory:
\  $<\OO(x) \OO(y) \OO(z)> \sim
\d (x-y) {1 \over |y-z|^8} + \d (x-z) {1 \over |y-x|^8} + \d (y-z) {1 \over |x-z|^8}$.
This is also suggested  by the field redefinition argument in 
 \fiedr.}
Given that in $\p C \p C$  case the  graviton exchange contributes only  to 
the $t$-channel,  one  would be   led to  the  conclusion 
that the  scalar and graviton exchange amplitudes 
should be proportional to each other (up to  possible  subtractions 
mentioned above).
That would, in turn, imply that  all these massless   scalar 
four-point  
amplitudes should be vanishing.
Given these  somewhat surprising conclusions, 
it remains to be seen if the crossing  symmetry can  actually   
 be  realised 
in  a  four-dimensional  CFT.

\appendix{A} {Scalar propagator  in $AdS_{d+1}$}

As in \witt\ we  consider the  Anti de Sitter space of dimension $D=d+1$ with 
Euclidean signature and the (half-space) metric  
\eqn\adsm{
ds^{2} = g_{0\m\n} dx^\m dx^\n= {1 \over x^2_0} (d x^2_0  + d x_{i}^{2}) \  .
}
The AdS$_{d+1}$  bulk  indices  will be 
denoted by $\m,\n, ...$ and will take values $ 0, 1,...,  d$.
We shall  use the notation  $x= (x_0, \vex)$, 
$\vex = (x_i)$, \ 
$i = 1, ..., d$.

Considering  the  Euclidean  action for a massive scalar 
$$
I  = \ha \int d^{d+1} x \sqrt{g_0}\, [ (\del_\m \p)^{2} + m^{2} \p^{2}]\ , 
$$
we are to solve 
$$
(D^{2} - m^{2}) G(x,y) = - {1 \over \sqrt{g_0}} \d (x-y)\ . 
$$
More explicitly, in the coordinate system \adsm, this equation becomes:
$$
x_0^n \del_{\m} [ x_0^{-n+2} \del_{\m} G(x,y) ] -m^2 G(x,y)
= - \d (\vec{x}-\vec{y}) \d (x_0 - y_0) y_0^n\ . 
$$
Let  $G(x,y) = x_0^{d/2} H(x,y)$.
Then 
\eqn\green{
(\De_{\n} + \del_i^2) H (x,y) = - y_0^{{d-2 \over 2}}
\d (\vec{x}-\vec{y}) \d (x_0 - y_0)\ , 
}
where $\n^2 = m^2 + {1 \over 4}d^2$ and the operator 
 $\De_\n$  defined by 
$$
\De_\n = \del_0^2 + {1 \over x_0} \del_0 - {\n^2 \over x_0^2}\ ,
\,\,\,\,\,\,$$
has Bessel functions as its eigenvalues
$$
\De_{\n} J_{\n} (w x_0) = - w^2 J_{\n} (w x_0)\ . 
$$
The delta-functions can be written in terms of orthonormal functions:
$$
\d (x_0 - y_0) = y_0 \int_{0}^{\inf} \! dw \, w J_{\n} (w x_0) 
J_{\n} (w y_0)
, \,\,\,\,\,\, \ \ \ \ \ 
\int {d^d  {k} \over (2 \pi)^{d}} e^{i \vec{k} \cdot (\vec{x}-\vec{y})}
= \d (\vec{x}-\vec{y})\ .  
$$
We expand the Green function in a similar fashion:
$$
H(x,y) = \int {d^d  {k} \over (2 \pi)^{d}} 
e^{i \vec{k} \cdot \vec{x}} 
\int_{0}^{\inf} \! dw \, w J_{\n} (w x_0)
\tilde{H}(w,k;y)\ . 
$$
Then the substitution into \green\ leads to
$$
\tilde{H}(w,k;y) = y_0^{d/2} { J_{\n} (w y_0) \over w^2 + k^2}
e^{- i \vec{k} \cdot \vec{y}} \ .
$$
The scalar Green function in 
the  Anti-de Sitter space \adsm\ is thus given by  
$$
G(x,y) = (x_0 y_0)^{d/2} \int {d^d  {k} \over (2 \pi)^{d}} 
\int_{0}^{\inf} dw\  w  {1 \over w^2 + k^2} e^{i \vec{k} \cdot (\vec{x}-\vec{y})}
J_{\n} (w x_0) J_{\n} (w y_0)
$$ 
\eqn\spro{
=  (x_0 y_0)^{d/2} \int {d^d  {k} \over (2 \pi)^{d}} \, 
e^{i \vec{k} \cdot (\vec{x}-\vec{y})} I_\n (k x_0^<) K_\n (k x_0^>)
\ , }
where $x_0^<$($ x_0^>$) is the smaller (larger) number 
among  $x_0$ and $y_0$.

\appendix{B} {Scalar contribution to four-point functions}

The contribution of a scalar operator $\P$ of dimension 
$\l$  to the CPWE of a four-point function of scalar
operators $<0|\P_{1}(x_1)\P_{2}(x_2)\P_{3}(x_3) \P_{4}(x_4)|0>$
has been worked out long time ago in \ferr. 
The same expression should also correspond to the 
scattering amplitude in AdS \fourt. 
Here we shall quote the result of \ferr.   
We will use the following definitions: \ 
$\S_{ij} = \l_i + \l_j$,\  $\De_{ij} = \l_i - \l_j$,\  $x_{ij}
= |x_i-x_j|$,\  and
$$
\rho = {x_{14}^2 x_{23}^2 \ov x_{12}^2 x_{34}^2}, \,\,\,\,\,\,\,\,\ \
\ \ 
\eta = {x_{13}^2 x_{24}^2 \ov x_{12}^2 x_{34}^2} \ , 
$$
where $\l_i,\  i=1,2,3,4$ are the conformal 
dimensions  of $\P_i$.
Then 
$$
<0|\P_{1}(x_1)\P_{2}(x_2)\P_{3}(x_3) \P_{4}(x_4)|0>
= x_{12}^{\S_{12}-\De_{12}} x_{13}^{-\De_{12}-\De_{34}}
x_{14}^{\De_{34}-\De_{12}} x_{34}^{\De_{12}-\S_{34}}
f_0(\rho, \eta) \ , 
$$
with
$$
f_0 (\rho, \eta) = c_1 \eta^{-\ha (\l - \De_{12})} \bigg [
F_{4} \big ( \ha (\l + \De_{34}), \ha (\l - \De_{12}),
\l +1-\ha d, \ha (\De_{34}-\De_{12})+1; {1 \ov \eta},
{\r \ov \e} \big) 
$$
$$
+
c_2 ({\r \ov \e})^{-\ha (\l -\De_{12})}
F_{4} \big ( \ha (\l - \De_{34}), \ha (\l - \De_{12}),
\l +1-\ha d, 1- \ha (\De_{34}+ \De_{12}); {1 \ov \r},
{\e \ov \r} \big) \bigg]
$$
When $\De_{12} = \De_{34}=0$, this  simplifies into
$$
f_0 (\rho, \eta) = c'_1 \eta^{-\ha \l} \bigg [
F_{4} \big ( \ha \l, \ha \l,
\l +1-\ha d, 1; {1 \ov \eta},
{\r \ov \e} \big) 
$$
$$+\ 
c'_2 ({\r \ov \e})^{-\ha \l}
F_{4} \big ( \ha \l, \ha \l,
\l +1-\ha d, 1; {1 \ov \r},
{\e \ov \r} \big) \bigg] \ . 
$$
In above $c_1, c_2, c'_1, c'_2$ are some numerical constants and 
$F_4$ is a double hypergeometric function.

\bigskip\bigskip

\centerline {\ \bf Acknowledgments}
\bigskip
We acknowledge the support 
 of PPARC and  of  the European
Commission TMR programme grant ERBFMRX-CT96-0045.
We  are  grateful  to  S. Ferrara, D. Freedman,  I.  Klebanov, 
R. Metsaev and A. Ritz
for stimulating  discussions. 

\listrefs
\bye